\documentclass[%
aps,
jmp,%
amsmath,amssymb,
reprint,%
]{revtex4-2}

\usepackage{graphicx}
\usepackage{dcolumn}
\usepackage{bm}
\usepackage{color}
\usepackage{CJK}
\usepackage{fancyhdr}
\usepackage{calc}
\usepackage{float}
\usepackage{mpgmpar} 
\usepackage{amsmath,amsthm} 
\usepackage{bm,here}
\usepackage{lipsum}
\usepackage{booktabs}

\begin{document}

\preprint{AIP/123-QED}
\title[Physical Review Applied]{Origin of Enhanced Electromechanical Coupling in (Yb,Al)N Nitride Alloys}

\author{Junjun Jia}
\email{jia@aoni.waseda.jp}
\affiliation{%
Global Center for Science and Engineering (GCSE), Faculty of Science and Engineering, Waseda University, 3--4--1 Okubo, Shinjuku, Tokyo 169--8555, Japan.
}%

\author{Takahiko Yanagitani}
\email{yanagitani@waseda.jp}
\affiliation{%
Graduate School of Advanced Science and Engineering, Waseda University, 3--4--1 Okubo, Shinjuku, Tokyo 169-8555, Japan 
}%
\affiliation{%
Kagami Memorial Research Institute for Materials Science and Technology, Waseda University, 2--8--26 Nishiwaseda, Tokyo 169--0051, Japan 
}%
\affiliation{JST PRESTO, 4--1--8 Honcho, Kawaguchi, Saitama 332-0012, Japan.}

\date{\today}

\begin{abstract}
Our experiments demonstrate that alloying the cubic--phase YbN into the wurtzite--phase AlN results in clear mechanical softening and enhanced electromechanical coupling of AlN. First--principle calculations reproduce experimental results well, and predict a maximum 270\% increase in electromechanical coupling coefficient caused by (1) an enhanced piezoelectric response induced by the local strain of Yb ions and (2) a structural flexibility of the (Yb,Al)N alloy. Extensive calculations suggest that the substitutional neighbor Yb--Yb pairs in wurtzite AlN are energetically stable along $c$ axis, and avoid forming on the basal plane of wurtzite structure due to the repulsion between them, which explains that (Yb,Al)N films with high Yb concentrations are difficult to fabricate in our sputtering experiments. Moreover, the neighbor Yb--Yb pair interactions also promote structural flexibility of (Yb,Al)N, and are considered a cause for mechanical softening of (Yb,Al)N. 

\end{abstract}

\maketitle

\section{Introduction}
AlN films have attracted considerable research interest for applications in electroacoustic devices, e.g., AlN film bulk acoustic resonator (FBAR) in microwave communication, because of their high thermal stability, and high operating frequencies, high Q factor, and high reliability.\cite{Muller2009, Yanagitani2014, Yokoyama2012} A major drawback is their low electromechanical coupling coefficient $k^2_t$, which is considerably smaller than those of perovskite--based oxides,\cite{Yanagitani2014, Ristic1983} which leads to loss in the microwave and low--bandwidth filters. The common experimental approach to improve $k^2_t$ is to alloy AlN with rocksalt nitrides (ScN, YN, and Ti$_{0.5}$Mg$_{0.5}$N).\cite{Akiyama2009, Akiyama2009-2, Tholander2013, Tholander2015}  A successful example is AlN alloying with rocksalt structured ScN, donated as (Sc,Al)N (Sc$_x$Al$_{1-x}$N), which shows a considerable enhancement in $k^2_t$ at $x$=0.43.\cite{Umeda2013} 

From the perspective of materials design, when alloying with rocksalt nitrides, a giant enhancement in $k^2_t$ can be considered as (1) an increase in the field--induced strain $via$ the increase in the longitudinal piezoelectric constant $e_{33}$, and (2) the simultaneous decrease in the longitudinal elastic stiffness $c_{33}$ based on $k_{t}^2 \sim e_{33}^2/\epsilon_{33} c_{33}$, where $\epsilon_{33}$ is the dielectric constant.\cite{Feneberg2007} In the (Sc,Al)N alloy, the enhancement in $k^2_t$ can be attributed to the giant piezoelectric response and mechanical softening due to the structural frustration, when the rocksalt endmember ScN is added to the alloy near phase boundary ($\sim$50\%) between the wurtzite and cubic phases.\cite{Manna2018} In spite of the giant piezoelectric response near phase boundary, the stability of the wurtzite phase at such high Sc concentrations is rather poor due to thermodynamic driving forces for phase separation, which leads to destruction of the piezoelectricity.\cite{Hoglund2010} To improve the stability of the wurtzite phase during the fabrication, a feasible approach is to discover the alloy system with the low wurtzite--to--rocksalt transition concentration, which is expected to fabricate more easily.\cite{Manna2018}  Likewise, as an indication for mechanical softening, the elastic constant $c_{33}$ for (Sc,Al)N decreases from 342 to 130 GPa with Sc dopant increasing from 0\% to 50\%.\cite{Tholander2013} The significant softening near phase boundary causes a reduction in the common figure of merit of the resonator,\cite{Syms2005} and thus makes the material less attractive for practical applications.\cite{Manna2017}  

Alloying AlN with transition--metal nitrides as alternative materials to (Sc,Al)N have been explored to investigate. Theoretical calculations indicate that alloying AlN with CrN [(Cr,Al)N] or YN [(Y,Al)N] produces comparable $e_{33}$ and $c_{33}$ with (Sc,Al)N, and is expected to generate a high $k^2_t$. However, so far, no experimental evidence (including our experiments) shows that their $k^2_t$ values increase with increasing alloying element concentration as (Sc,Al)N does. In particular, an opposite behavior for (Cr,Al)N is observed, i.e., the $k^2_t$ values decrease with increasing CrN concentration, which is possibly due to phase transition from a piezoelectric to a non-piezoelectric crystal phase.\cite{Suzuki2019} Therefore, important aspects of electromechanical coupling in AlN--based piezoelectric materials require both experimental investigation and theoretical calculation.

In order to achieve the high $k^2_t$ value with low alloying concentration, a reasonable approach is to substitutionally dope AlN by using the impurity element with larger ionic radius and different electronic configuration. Following the abovementioned scenario, alloying AlN with YbN, designated as (Yb,Al)N (Yb$_x$Al$_{1-x}$N),  is expected to cause a similar structural frustration to produce the mechanical softening, since YbN also has a rocksalt cubic structure (Fm$\bar{3}$m, 225), where Yb$^{3+}$ is bonded to six equivalent N$^{3-}$ ions and form a mixture of corner-- and edge--sharing YbN$_6$ octahedra. Nevertheless, Yb has a [Xe]4$f^{14}$$6s^{2}$ electronic configuration different from [Ne]3$s^{2}$$3p^{1}$ of Al, and has a larger atomic radius than the reported transition metal nitride endmembers (ScN and CrN).\cite{Akiyama2009, Akiyama2009-2, Tholander2013} These characteristics are considered to generate the resistance to plastic deformation and thus mitigate the excessive softening. Moreover, a larger atomic radius would be expected to produce a large internal strain, which possibly induces an increase in $e_{33}$. 

In this study, we investigate electromechanical coupling phenomenon in (Yb,Al)N from both experimental and theoretical aspects, and demonstrate that alloying the cubic YbN into the wurtzite AlN leads to clear mechanical softening and enhanced $e_{33}$, which improves $k^2_t$ up to $\sim$10\%, which is close to the reported value of 15.5\% for (Sc,Al)N.\cite{Umeda2013} We also provided a fundamental understanding of such dramatic enhancement in electromechanical coupling for (Yb,Al)N. 

\section{Experimental and Computational Details}

The combinatorial synthesis of $c$--axis oriented (Yb,Al)N films are fabricated by sputtering deposition. (Yb,Al)N films are deposited on Ti electrode film/silica glass substrate, where a highly oriented (0001) Ti electrode film is used as the under-layer electrode to form the high--overtone bulk acoustic resonator (HBAR) structure. The film structure is characterized by X--ray diffraction (XRD) with a $2\theta$--$\omega$ configuration (X'Pert PRO, PANalytical). The Yb concentrations are analyzed by electron probe microanalysis (EPMA) (JXA-8230, JEOL). The $k_t^2$ values and the film density are determined by comparing the experimental and theoretical longitudinal wave conversion loss curves {\it vs} the frequencies of HBAR resonators.\cite{Yanagitani2014, Iwata2020}  

For theoretical calculations, the structures representing hexagonal (Yb,Al)N alloys up to $x$=0.43 were first relaxed using density functional theory. The Vienna Ab--initio Simulation Package (VASP)\cite{Kresse1996} was used for structural optimization, using the generalised gradient approximation (GGA) as parameterized by Perdew $et$ $al.$ for the exchange-correlation potential.\cite{PBE} A 3 $\times$ 3 $\times$ 2 supercell was selected, and the special quasirandom structure (SQS)\cite{Zunger1990, Walle2013, Chen2011} method is used to model the random distribution of Yb in the wurtzite (Yb,Al)N structure with different $x$ values. The SQSs were generated by optimizing the locations of Yb atoms to minimize the Warren--Cowley pair short range order parameters, which were calculated up to the sixth coordination shell. Moreover, to realistically simulate the chemical disorder of actual alloy nitrides, we constructed $\sim$10 SQS structures with the same Yb concentration, and selected the SQS structure with the minimum total formation energy to calculate piezoelectric and elastic tensors. The Monkhorst--Pack $k$--point grids were set to 3 $\times$ 3 $\times$ 3. The plane wave basis set with a cutoff energy of 600 eV is adopted, and the total
energies were converged to less than 10$^{-9}$ eV for structural relaxation. The elastic and piezoelectric tensors were calculated using density--functional perturbation theory.\cite{Smith1993, Vanderbilt2000, Gonze1997, Wu2005}  

\section{Results and Discussion}

\subsection{Structural Characteristics}

All fabricated (Yb,Al)N films are oriented along the $c$--axis with various rocking curve full width at half maximum (FWHM), as shown in Fig. \ref{k33-V33}. No phases from Yb or YbN can be found in the XRD pattern.

\begin{figure}[h]
\begin{center}
\includegraphics[width=8cm]{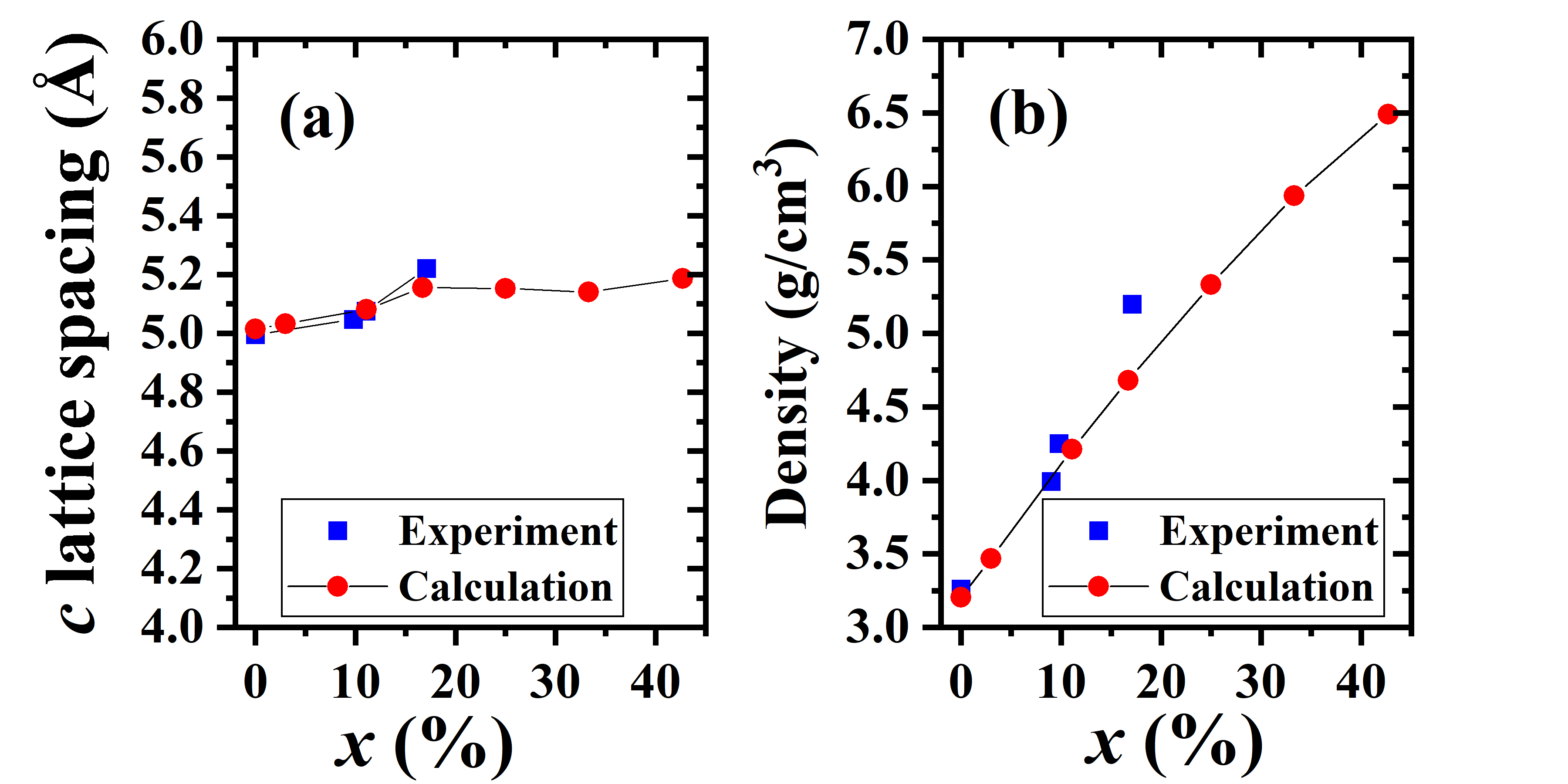}
\caption{The $c$ lattice spacing (a) and density (b)  of (Yb,Al)N films from experimental measurement and theoretical calculation. The solid square represents the measurement values, and the solid circle shows the calculated values. Note that (Yb,Al)N films with high Yb concentrations ($x>$0.25) are difficult to fabricate by sputtering.}
\label{lattice}
\end{center}
\end{figure}

The calculated lattice parameters of wurtzite AlN are $a$=3.128 $\rm \AA$ and $c$=5.015 $\rm \AA$, which are close to reported experimental and theoretical values.\cite{Damme1989, Bernardini1997} Table \ref{Structure} shows the calculated lattice constants of (Yb,Al)N. Both $a$-- and $c$--axis lattice spacings monotonically increase with increasing Yb concentration due to the substitution of Yb$^{3+}$ ion (ionic radius: 0.87 \AA) into the host Al ion (ionic radius: 0.54 \AA). Whereas, XRD patterns show a similar tendency, where the (0002) plane gradually shifts toward the low--angle side, demonstrating that Yb atoms are substituted into host Al sites during the nucleation and/or growth. The extrapolated $c$--lattice constants from the precise XRD measurements are consistent with those from our first--principle calculations, as shown in Fig. \ref{lattice}. Moreover, our calculations show a decreasing trend in the $c/a$ ratio, which suggests that an increase in lattice spacing in the $a$ axis is larger than that along the $c$ axis.

\begin{table}[h]
\caption{\label{Structure} Calculated structural parameters of (Yb,Al)N. }
\begin{ruledtabular}
\begin{tabular}{cllcccccccc}
        & $a$ (\AA) & $c$ (\AA) &$c/a$  & $\rho$ ($g/cm^{3}$)\\
\hline
AlN          & 3.128                        &5.015   &1.603  &3.203  \\
$x$=0.03   & 3.146$\pm$0.000   &5.032$\pm$0.000   &1.600  &3.469  \\ 
$x$=0.11   & 3.201$\pm$0.001   &5.081$\pm$0.002   &1.596  &4.211  \\
$x$=0.17   & 3.234$\pm$0.005   &5.156$\pm$0.004   &1.602  &4.679   \\
$x$=0.25   & 3.295$\pm$0.007   &5.153$\pm$0.016   &1.568  &5.332   \\
$x$=0.33   & 3.357$\pm$0.024   &5.141$\pm$0.010   &1.523  &5.935   \\
$x$=0.43   & 3.408$\pm$0.006   &5.187$\pm$0.014   &1.523  &6.490   \\
\end{tabular} 
\end{ruledtabular}
\end{table}

We next inquire why the $c/a$ ratio decreases as $x$ increases. Let's start from the question: what are the energetics involved in finding two substitutional Yb atoms in the wurtzite AlN structure? From this aspect, we calculated the pair interaction energy $\Delta E^{(n)}$ as the difference between the total energy of two $n$th neighbor Yb atoms and for two Yb atoms at infinite separation (twice the energy of an isolated Yb atom),\cite{Zunger1999} i.e.,

\begin{equation}
\begin{split}
  \Delta E^{(n)}=\left[E(\mathrm{Al}_{(\frac{m}{2}-2)}\mathrm{Yb}_2\mathrm{N}_{\frac{m}{2}}) + E(\mathrm{Al}_{\frac{m}{2}}\mathrm{N}_{\frac{m}{2}})  \right] \\
-2E(\mathrm{Al}_{(\frac{m}{2}-1)}\mathrm{Yb}\mathrm{N}_{\frac{m}{2}}),
\label{pair}
\end{split}
\end{equation}

where $n$ denotes the pair index, and $m$=72 is the number of atoms in the supercell. The energies correspond to the fully relaxed supercell calculations. In the calculations, an identical $k$--mesh (3 $\times$ 3 $\times$ 3) is applied for all three total energies for consistency. The lattice parameter for all supercells has been chosen to correspond to the theoretical AlN lattice constants.

\begin{figure}[t]
\begin{center}
\includegraphics[width=8 cm]{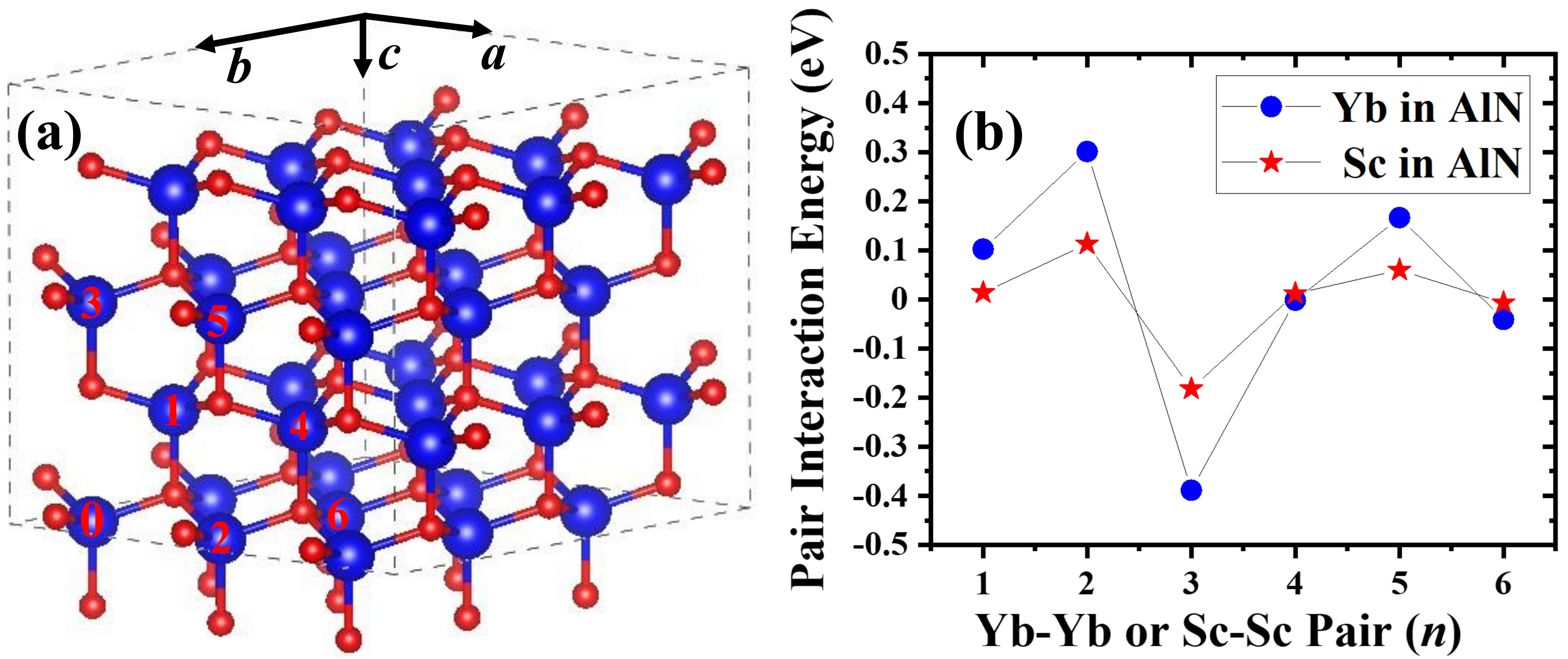}
\caption{(a) the six pair configurations in the 3 $\times$ 3 $\times$ 2 supercell (Blue: Al, and Red: N atoms). A Yb--Yb pair ($n$) is defined as the two $n$th neighbor substitutional Yb atoms relative to a Yb atom located at the origin ``O". (b) The pair interaction energies $\Delta E^{(n)}$ for $n$=1, 2, 3, 4, 5, 6th neighbor Yb pairs. For comparison, $\Delta E^{(n)}$ in Sc doped AlN was also calculated.}
\label{pair-fig}
\end{center}
\end{figure}

The $\Delta E^{(n)}$ is shown in Fig. \ref{pair-fig}. The result for $\Delta E^{(n=3)}$ is negative (--0.38 eV), indicating that the binding between the Yb atoms along $c$ axis is energetically stable. $\Delta E^{(n=4)}$ and $\Delta E^{(n=6)}$ are close to zero, but remains negative. The positive $\Delta E^{(n)}$ for $n$=1, 2, and 5 indicates the repulsive interaction between the Yb atoms at these positions. Here, the attractive pair interaction is considered to induce the compressive stress along the $c$ axis ($n$=3), and the repulsive pair interaction ($n$=2) is expected to give rise to the tensile stress along the basal plane. Thus, their superposition leads to a decreasing $c/a$ ratio with increasing $x$. Note that the calculation results for $\Delta E^{(n)}$ using a bigger supercell (5 $\times$ 5 $\times$ 3) with $a$=15.64 \AA\  and $c$=15.04 \AA\  show the same tendency, suggesting that $\Delta E^{(n)}$ is an intrinsic physical parameter for the (Yb,Al)N alloy.

Moreover, the pair interaction energies would suggest that Yb--Yb pairing avoids $n$=1, 2, and 5th neighbor shells, and favors $n$=3, 4 and 6th neighbor configurations, especially $n$=3. $\Delta E^{(n)}$ in (Sc,Al)N shows the similar trends but has smaller value. For a growth process under thermal equilibrium, the Yb atoms should most likely pair into the third nearest neighbor configurations (namely Yb--Yb pairing along $c$ axis) rather than be randomly distributed. This also explains our experimental observations that (Yb,Al)N films with high Yb concentrations ($x>$0.25) are difficult to fabricate in sputtering experiments because the repulsion among Yb dopants in the basal plane suppresses Yb substitution into Al sites. A similar repulsion among dopants is also present in Sn--doped In$_2$O$_3$. \cite{Yamada1999, Jia2019, Jia2021}

Based on the above results, a SQS cell with a random distribution of Yb atoms cannot model the spatial configuration of Yb atoms in wurtzite AlN. In this study, we constructed $\sim$10 different SQS structures with the same Yb concentration, and selected the SQS structure with the minimum total formation energy to calculate its piezoelectric and elastic tensors.

\subsection{Electromechanical Coupling}

Since obtaining accurate measurement of $c_{33}$ and $e_{33}$ along $c$ direction for nitride alloy films is quite difficult, especially for the thin films with a thickness of several hundred nanometers, in this study, the experimentally measured $k_{t}^2$ and acoustic wave velocity $V$ are directly compared with the theoretically calculated values.  

\begin{figure}[h]
\begin{center}
\includegraphics[width=8 cm]{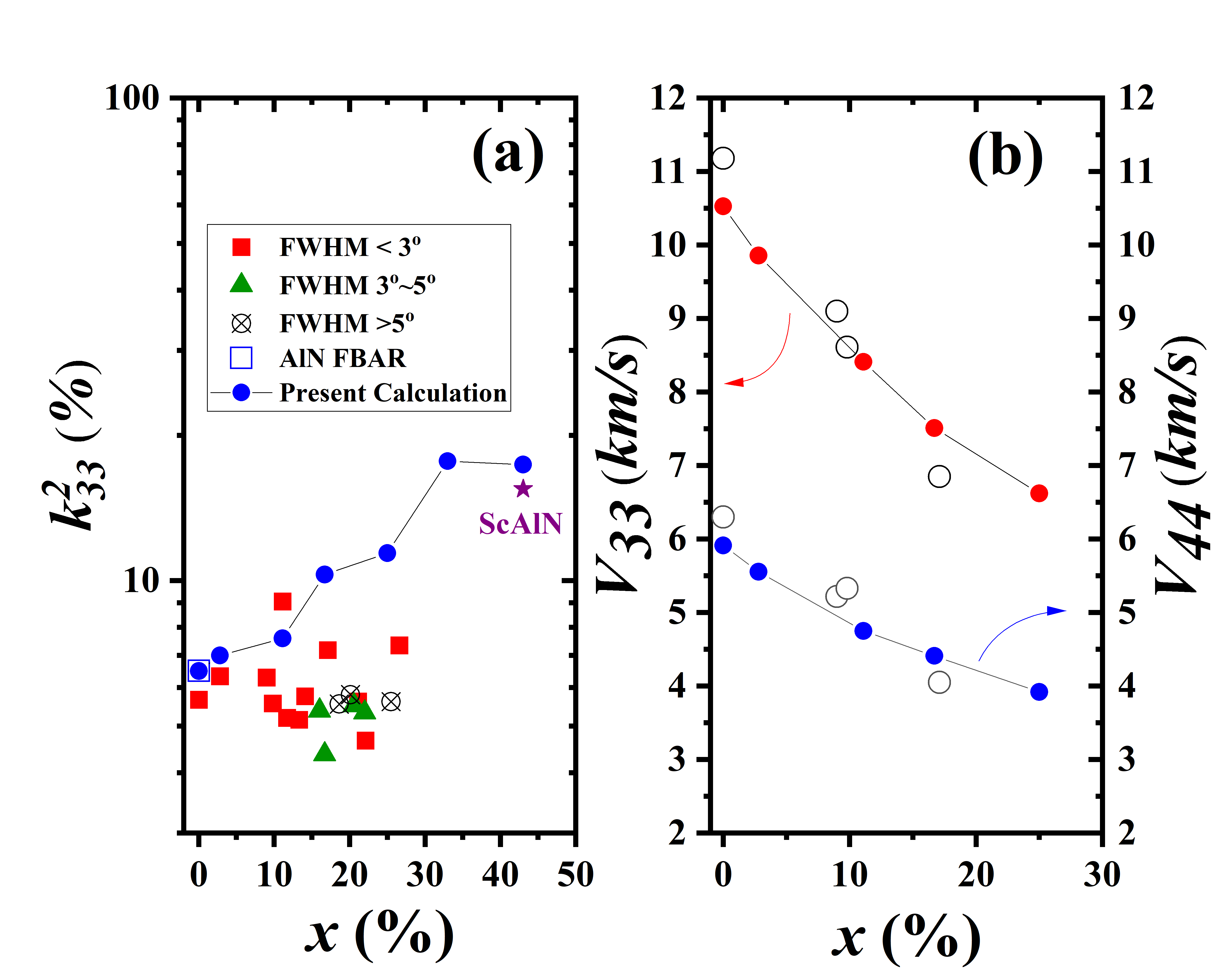}
\caption{(a) The calculated $k_{33}^2$ values are compared with the experimentally determined $k_t^2$ for (Yb,Al)N films with different rocking curve FWHMs. The smaller is the FWHM, the stronger is the $c$--axis orientation. (b) The calculated $V_{33}$ and $V_{44}$ values are compared with the experimentally measured $V_{33}^D$ and $V_{44}^E$ for (Yb,Al)N films with FWHM$<$3$^\circ$, respectively. Connected lines represent the calculated values, whereas the scattered points show the experimentally measured ones. In Fig. (a), the symbol $\square$ represents the measurement value for the general pure AlN FBAR with Mo electrodes, and the asterisk shows the $k_{t}^2$ value for (Sc,Al)N at $x$=43\%.\cite{Umeda2013}}
\label{k33-V33}
\end{center}
\end{figure}

Because $k_{t}^2$ is known to be approximately equal to the longitudinal electromechanical coupling coefficient $k_{33}^2$ for the $c$-oriented AlN films,\cite{Feneberg2007} $k_{t}^2$ can be estimated by $k_{t}^2 \sim k_{33}^2=e_{33}^2/\epsilon_{33} c_{33}$. In Fig. \ref{k33-V33} (a), the measured $k^2_t$ values are compared with the calculated $k_{33}^2$ values. The calculated $k_{33}^2$ are in good agreement with the experimentally measured tendency for $k_{t}^2$ up to approximately $x$$\sim$0.11, which increases as Yb concentration increases. The (Yb,Al)N films fabricated at various conditions show a maximal $k_t^2$ value of 9.05\% at approximately $x$=0.11. In contrast, our theoretical calculations suggest the maximal $k^2_{33}$ of 17.6\% at $x$=0.33. At high Yb concentration ($x$$>$0.12), the calculated $k_{33}^2$ is larger than the measured $k_t^2$. This is most likely due to the fact that the experimentally determined $k_t^2$ often depends on the crystallinity of thin films and the crystal orientation, as well as the polarization direction. Moreover, as revealed by the calculations of the pair interaction energies, the repulsion between Yb--Yb pairs on the basal plane is considered to hinder further substitutional Yb doping.  

Acoustic wave velocity is also a measurable experimental quantity, and it can be precisely determined by acoustic wave resonance spectra.\cite{Ohashi2008} Under a constant external electric field, elastic constants ($c_{33}^E$ and $c_{44}^E$) can be experimentally determined by the following equations, 

\begin{equation}
c_{33}^E = \rho(V_{33}^D)^2(1-k_{33}^2)
\end{equation}

and 
\begin{equation}
c_{44}^E = \rho(V_{44}^E)^2,
\end{equation}

where $\rho$ is the density, and $V_{33}^D$ and $V_{44}^E$ are longitudinal and shear acoustic wave velocities, respectively. In Fig. \ref{k33-V33} (b), the calculated $V_{33}$ and $V_{44}$ values are consistent with experimentally measured $V_{33}^D$ and $V_{44}^E$ values, respectively. Such quantitative agreement between our calculation (0 K) and the experimental results measured at room temperature allows us to expect that acoustic phonon contribution to piezoelectricity in (Yb,Al)N alloys is small up to room temperature. 

\begin{table*}[ht]
\caption{\label{cij}
Elastic ($c_{ij}$) and piezoelectric ($e_{ij}$) constants of (Yb,Al)N with different Yb concentration $x$, where $c^{E}_{33}$ and $c^{E}_{44}$ are the experimental values by measuring (Yb,Al)N films with FWHM$<$3$^\circ$ in this study. Elastic constants are in GPa, and piezoelectric constants are in C/m$^2$. The calculated and measured $c_{ij}$ and $e_{ij}$ values for AlN from other groups are provided for comparison, and our calculations agree well with previous ones.\cite{Bernardini1997, Xie2012} Note that $\epsilon_{33}$ is the 33 component of the dielectric tensor.}
\begin{ruledtabular}
\begin{tabular}{lrlllllccccc}
AlN & $\epsilon_{33}$ & $c_{11}$ & $c_{12}$ & $c_{13}$ &  $c_{33}$  & $c_{44}$ & $c^{E}_{33}$ & $c^{E}_{44}$  & $e_{31}$ & $e_{33}$ & $k^2_{33} (\%)$ \\
\hline \\
This study \\
AlN            & 9.76  & 375  & 128 & 98   & 355 & 112& 380    &127   & -0.58  & 1.46 & 6.5\\
$x$=0.03    & 9.97  & 360  & 128 & 102 & 337 & 107 & --       & --       & -0.58  & 1.50  & 7.0\\
$x$=0.11    & 11.16  & 312  & 124 & 105 & 298 & 95 & 302     &109   & -0.63  & 1.56  & 7.6\\
$x$=0.17    & 10.99  & 294  & 129 & 113 & 264 & 91 & 207     & 78   & -0.59  & 1.72 &10.3\\
$x$=0.25    & 11.67  & 276  & 123 & 116 & 234 & 82 & --          & --    & -0.59  & 1.76 & 11.4\\
$x$=0.33    & 12.84  & 255  & 139 & 119 & 192 & 85 & --       & --    & -0.68  & 2.17 & 17.7\\
$x$=0.43  & 13.15  & 231  & 126 & 117 & 182 & 77   & --          & --    & -0.65  & 2.12 & 17.4\\

References for AlN \\
{\it Measured by Tsubouchi et al.}\footnotemark[1]  & -- & 345 & 125 & 120 & 395 & 118 & -- &--  & -- & --& --\\
{\it Measured by McNeil et al.}\footnotemark[2]  & --  & 411$\pm$10 & 149$\pm$10 & 99$\pm$4 & 389$\pm$10 & 125$\pm$5 & -- &--  & -- & --& --\\
{\it Calculated by Xie et al.}\footnotemark[3]  & --  & 376  & 126 & 98 & 356 & 116 & -- & --  & -- & --& --\\
{\it Calculated by Bernardini et al.}\footnotemark[4]  & --  & -- & -- & -- & -- & -- & --& -- &-0.60  & 1.46 & --\\
\end{tabular}
\end{ruledtabular}
\footnotetext[1]{Reference.~\cite{Tsubouchi1981}.}
\footnotetext[2]{Reference.~\cite{McNeil1993}.}
\footnotetext[3]{Reference.~\cite{Xie2012}.}
\footnotetext[4]{Reference.~\cite{Bernardini1997}.}
\end{table*}

Based on the quantitative agreement between our calculation and experimentally measured results, we investigated the microscopic mechanism for the $k^2_{t}$ change with Yb concentration. The calculated dielectric constants, piezoelectric constants, and elastic moduli for (Yb,Al)N with different $x$ are listed in Table \ref{cij}. With an increase in the Yb concentration, $c_{33}$ linearly decreases up to $x$=33\%, and a slowdown is observed from $x$=33\% to 43\%. Our calculations reproduce the changing behavior of $c^E_{33}$ and $c^E_{44}$ well. Moreover, an increase in $e_{33}$ with the alloying Yb concentration is accompanied by a decrease in $c_{33}$; thus, the increasing piezoelectric response and mechanical softening are considered to cooperate to cause an increase in $k_{t}^2$ in (Yb,Al)N films.

\begin{figure}[h]
\begin{center}
\includegraphics[width=8 cm]{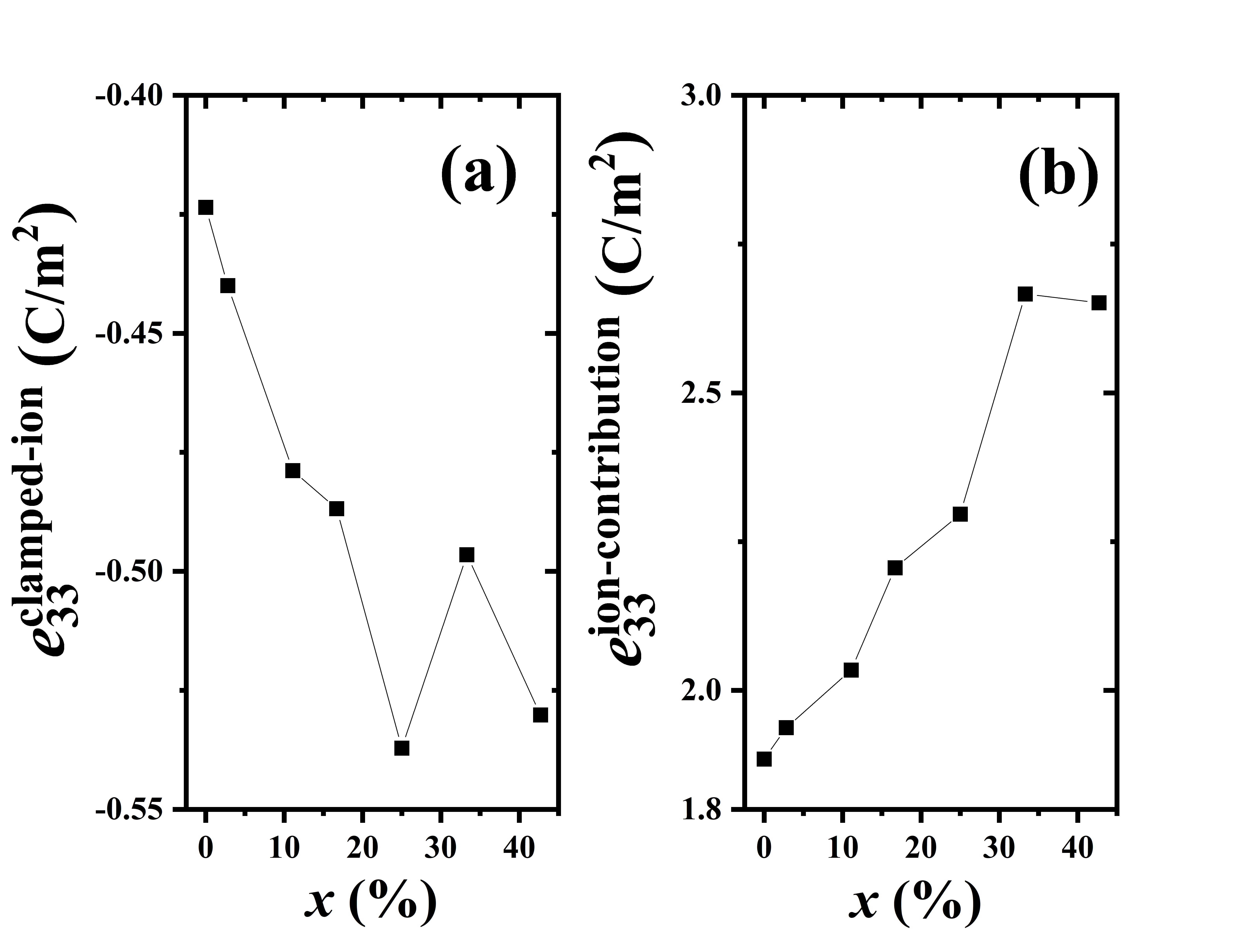}
\caption{Calculated contributions of $e_{33}$ introduced in Eq. (3). (a) the $clamped$--$ion$ contribution, and (b) the ionic contribution, which is related to the strain effect and the Born charges $Z_{33}$ calculated from Eq. (3).  }
\label{e33}
\end{center}
\end{figure}

From the microscopic perspective, $e_{33}$ can be divided into two main contributions: (1) an electronic response to the strain of the crystal structure ($clamped$--$ion$ term), and (2) the effect of internal strain on piezoelectric polarization.\cite{Tasnadi2010, Bernardini1997} Therefore, $e_{33}$ can be expressed as follows: 

\begin{equation}
  e_{33} (x) = e_{33}^{clamped-ion} (x) +\frac{4eZ_{33}(x)}{\sqrt{3}a(x)^2}\frac{du}{d\sigma},
\label{eqn-e33}
\end{equation}

where $e$ is the elementary charge, $a$ represents the equilibrium lattice parameter, $Z$ represents the dynamical Born or transverse charge in units of $e$, and $\sigma$ is the macroscopic applied strain. The wurtzite internal parameter is described by $u$, where the layered hexagonal phase differs from the wurtzite structure only in the internal parameter $u$ between metal and nitrogen sublattices ($u$=0.5 for the hexagonal phase, and $u$$\neq$0.5 for the wurtzite phase). In Fig. \ref{e33} (a), a monotonic decrease was observed for the $clamped-ion$ term up to $x$=0.25, followed by an increase up to --0.49 C/m$^2$ at $x$=0.33. In comparison, the ionic contribution for $e_{33}$, which represents the local structural sensitivity to macroscopic axial strain $\sigma$, shows a clear increase with increasing Yb concentration in Fig. \ref{e33} (b). Thus, an increased piezoelectric response should be mainly attributed to the internal structural strain.  

\begin{figure}[h]
\begin{center}
\includegraphics[width=8 cm]{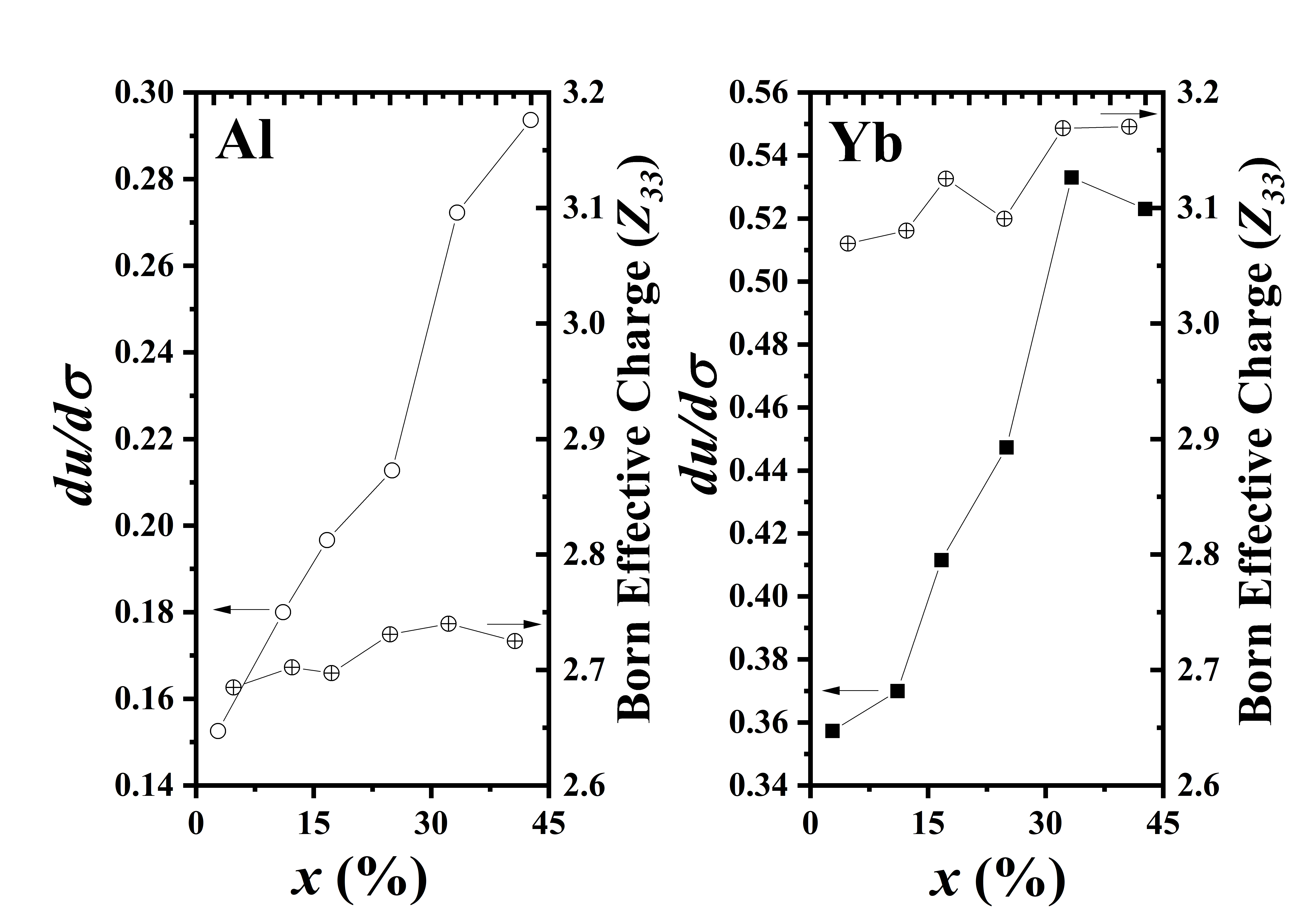}
\caption{Compositional weighted and site resolved internal strain sensitivity and Born effective charges of (Yb,Al)N.}
\label{strain}
\end{center}
\end{figure}

To understand the ionic contribution to the piezoelectricity of (Yb,Al)N, we calculated the compositional weighted and site--resolved internal strain sensitivity and dynamic Born charge. Fig. \ref{strain} shows that the internal strain sensitivity ($du/d\sigma$) around both Al and Yb sites increases as the Yb concentration increases, and the structural strain sensitivity around Yb sites is approximately two times larger than that around Al sites under different Yb concentrations. On the other hand, $Z_{33}$ for Yb ions remains at $\sim$3.1, and $Z_{33}$ for Al ions keeps at $\sim$2.7, which vary within approximately 10\% around the ionic nominal value of 3. A larger $Z_{33}$ indicates that the Yb ion has a large polarization response to the external electric field. The abovementioned comparison suggests that the change in the ionic contribution is mainly dominated by local structural distortion caused by alloying Yb.

\subsection{Mechanical Softening}

From the structural point of view, AlN has a wurtzite structure (6$mm$, 226), where Al$^{3+}$ is bonded to four equivalent N$^{3-}$ to form a corner-sharing tetrahedral structure, and YbN has a cubic structure (Fm$\bar{3}$m, 225) with a mixture of corner-- and edge--sharing octahedral structure YbN$_6$. The substitution of the Yb atom in Al sites is presumed to lead to structural flexibility, and then causes mechanical softening. Herein, we investigated the potential energy landscape of (Yb,Al)N at $x$=0.33, which, in theory, has the maximal $k_t^2$ value.

\begin{figure}[h]
\begin{center}
\includegraphics[width=9 cm]{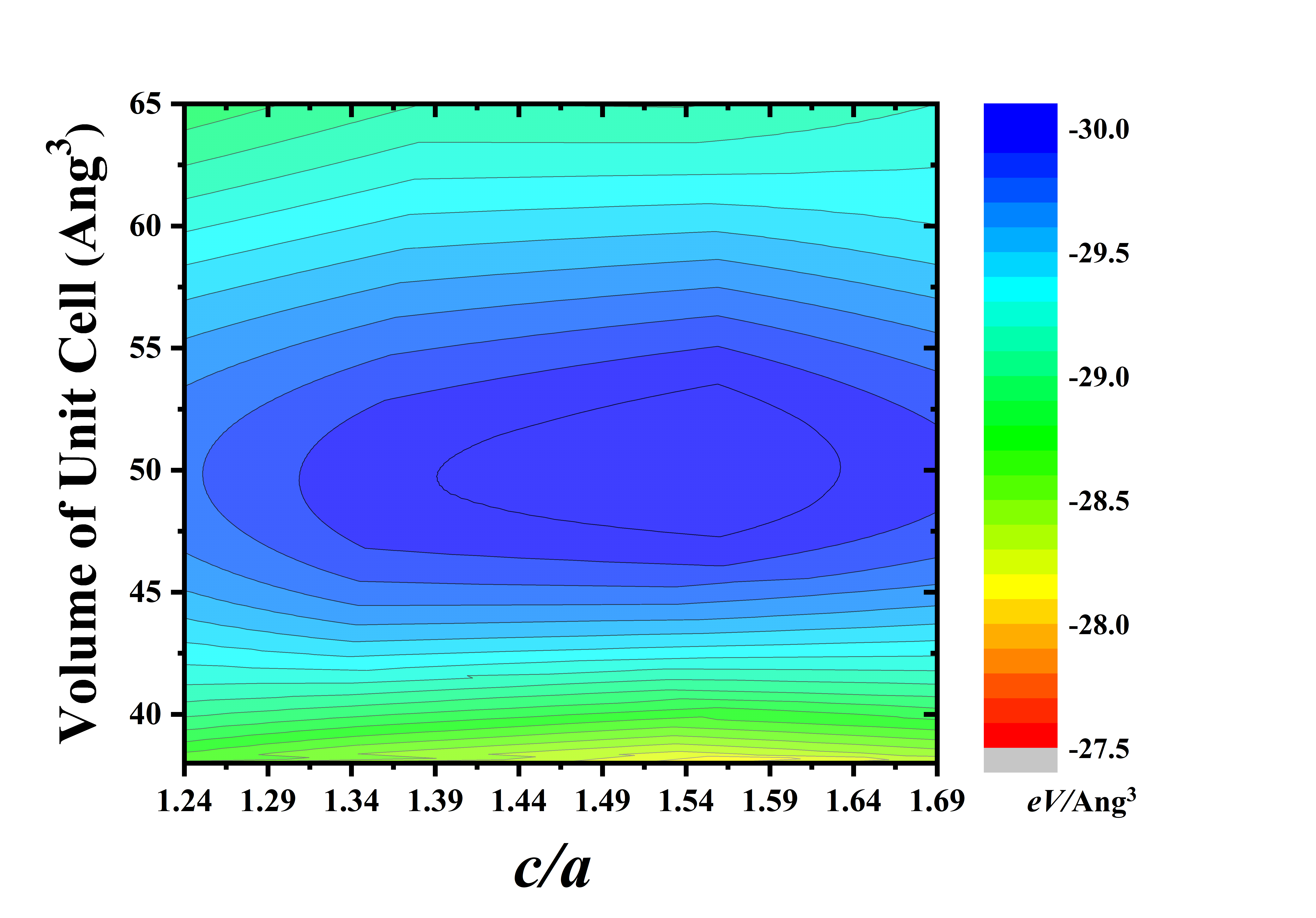}
\caption{Energy landscape of wurtzite (Yb,Al)N ($x$=0.33).}
\label{landscape}
\end{center}
\end{figure}

In Fig. \ref{landscape}, the potential energy landscape exhibits a flat, elongated structure, which is derived by the interpolation of energy points in the ($volume$, $c/a$ ratio) space. This landscape pattern differs from that of well-studied (Sc,Al)N.\cite{Tasnadi2010} In the (Sc,Al)N alloy system, the Sc atoms prefer to bind hexahedrally to nitrogen, and the resultant hexagonal ScN phase is energetically close to the wurtzite AlN phase, which acts as an energy saddle point.\cite{Bellaiche2002, Tasnadi2010}

\begin{figure}[h]
\begin{center}
\includegraphics[width=8 cm]{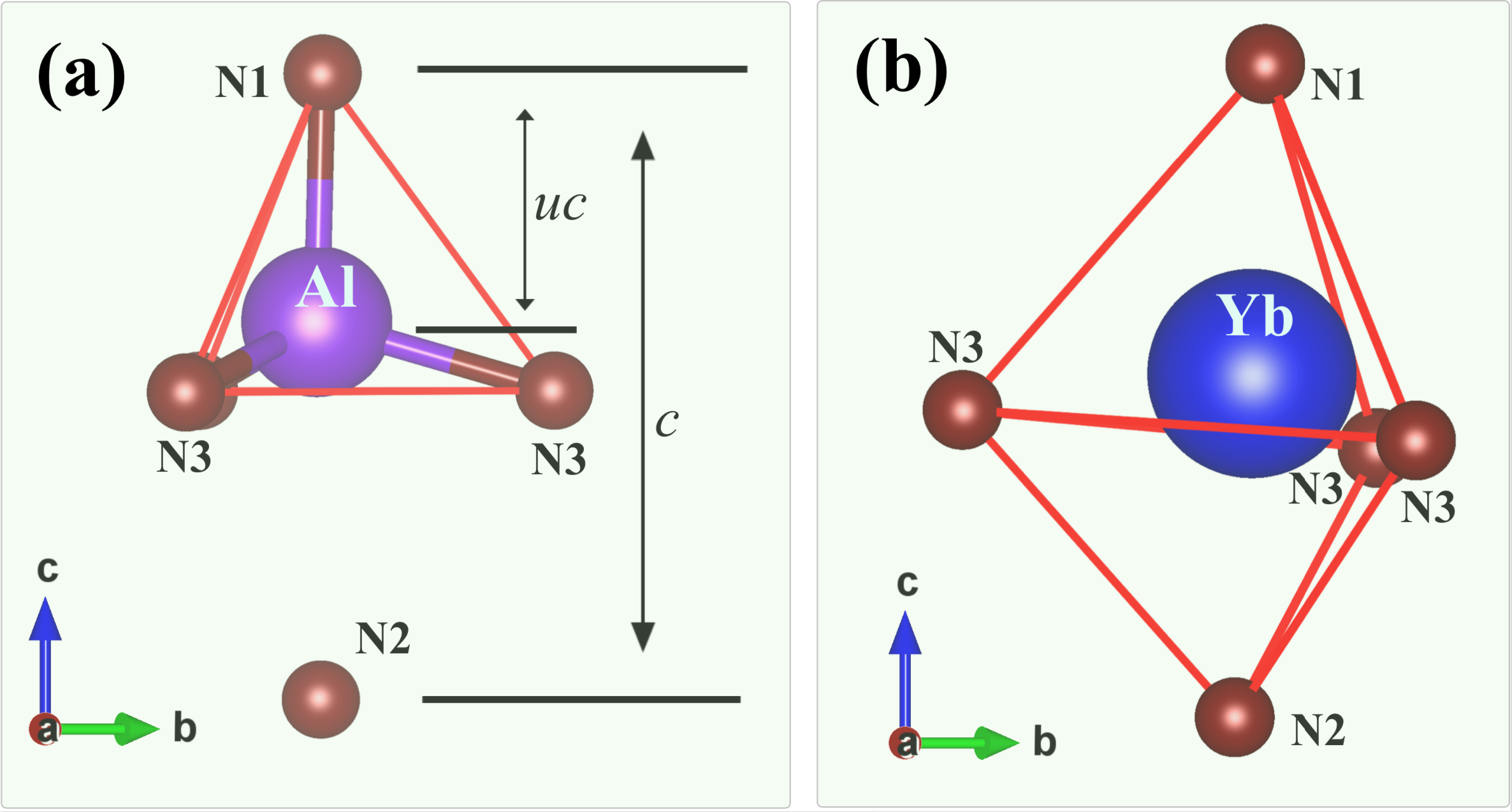}
\caption{Local structures of Yb and Al atoms in the relaxed (Yb,Al)N alloy system ($x$=0.33, and $c/a$=1.52). The index of nitrogen atoms in the local structure is given. The deformed bipyramidal structure in Fig. (b) is shown just for the eye-guide.}
\label{local}
\end{center}
\end{figure}

To reveal the mechanical softening mechanism in the (Yb,Al)N alloy system, we investigated the local structure of the Yb atom. In the relaxed (Yb,Al)N structure ($x$=0.33 and $c/a$=1.52), clear local structural distortion around Yb ion is observed, which differs from the local structure of the Al ion, as shown in Fig. \ref{local}. Depending on different local chemical environments, the Yb--N(1) bond length varies from 2.152 to 2.336 \AA\ with an average of 2.214 \AA, and the Yb--N(3) bond length varies from 2.073 to 2.279 \AA\ with an average of 2.172 \AA. Notably, Al--N tetrahedral bond lengths are 1.913 \AA\ in the parent wurtzite AlN structure, and the longer Yb--N(1) and Yb--N(3) bond lengths in the tetrahedral Yb$^{3+}$ structure are due to a larger ionic radius and more widely extended electronic orbitals of Yb ions (principal quantum number $n$=6). Moreover, the Yb--N(2) bond length varies from 2.540 to 3.249 \AA\ with an average of 2.963 \AA. The obvious difference between Yb--N(1) and Yb--N(2) ($u$$\neq$0.5) suggests (Yb,Al)N ($x$=0.33 ) maintains piezoelectric crystal phase. Furthermore, when the $c/a$ ratio is compressed to 1.40, which is around the edge of the lowest energy contour in Fig. \ref{landscape}, the shortest bond lengths of Yb--N(3) on the basal plane is 2.023 \AA, and the shortest bond lengths of Yb--N(1) is 2.142 \AA. The shortest bond length of Yb--N(2) is 2.356 \AA, comparable with the theoretical Yb--N bond length (2.464 \AA) in the cubic YbN structure. Instead of a YbN$_4$ tetrahedral structure, the local structure of Yb$^{3+}$ seems to have a deformed bipyramidal structure, which needs further investigation.

Based on the difference in the bond length between Yb--N(1) and Yb--N(2) under external pressure ($c/a$=1.40), a decrease in the $c/a$ ratio with increasing $x$ in Table \ref{Structure} cannot be connected to the phase transition from the wurtzite ($u$=0.5) to a quasi--stable layered hexagonal phase ($u$$\neq$0.5), as those in (Sc,Al)N.\cite{Tasnadi2010} Alternatively, the mechanical softening of (Yb,Al)N is attributed to structural flexibility, which is considered from the change of chemical bond type and/or orientation as well as the local structure after alloying with YbN, which flattens the energy potential landscape. Structural flexibility is beneficial to the change of wurtzite (Yb,Al)N toward a layered hexagonal phase under external pressure.

\begin{figure}[h]
\begin{center}
\includegraphics[width=8 cm]{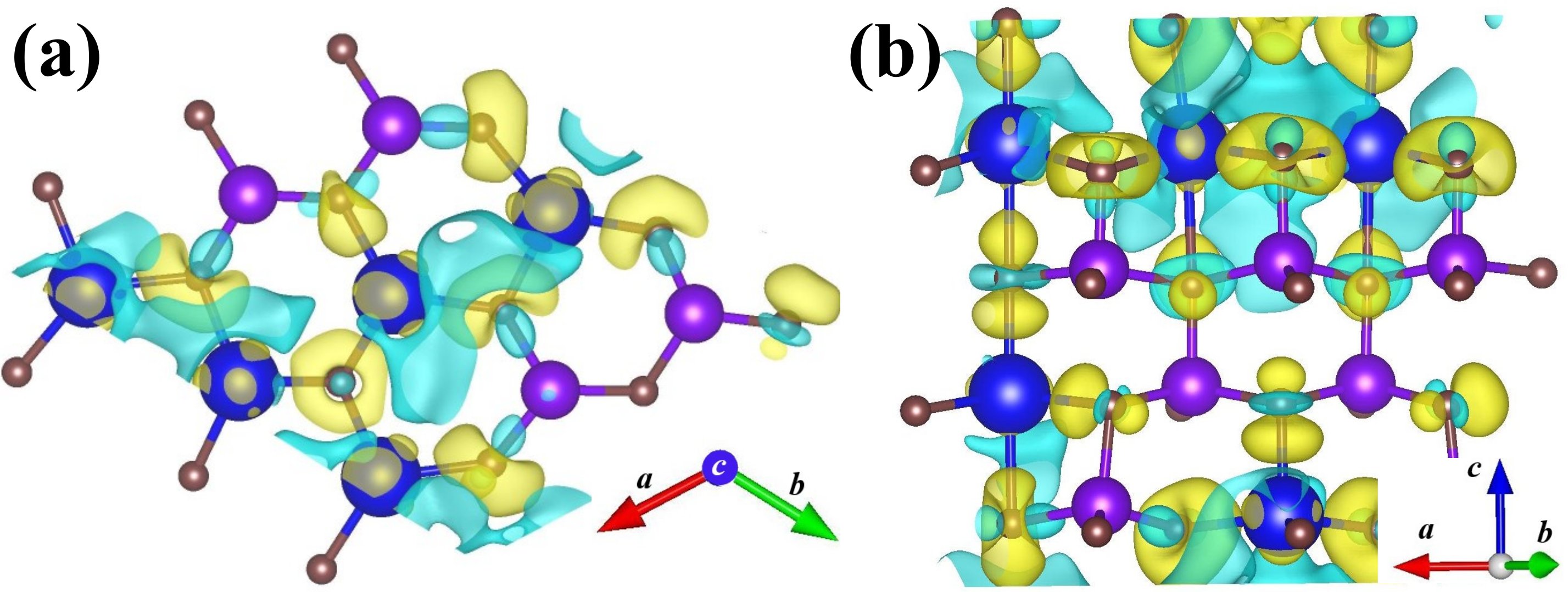}
\caption{Charge density differences of (Yb,Al)N with $x$=0.33 under the compressed state of $c/a$=1.40. (a) a $ab$ basal plane is selected, and  (b) a $ac$ plane is selected to show charge density difference along $c$ axis. The loss of electrons is indicated in light--blue color, and electron accumulation is indicated in yellow. Each isovalue is set to 10\% of the maximum. (Purple: Al, Blue: Yb, and Brown: N atoms)}
\label{CHG}
\end{center}
\end{figure}

To gain more microscopic insights about the mechanical softening, the charge density difference is calculated to study the chemical interaction among constituent atoms in (Yb,Al)N with $x=$0.33. We define the charge density difference as $\delta\rho$=$\rho_t$--$\rho_1$--$\rho_2$, where $\rho_t$, $\rho_1$, and $\rho_2$ are the charge densities of (Yb,Al)N, AlN, and isolated Yb atoms, respectively. In Fig. \ref{CHG} (a), the results of the redistribution of charge density obtained by $ab$ initio calculation show that there is considerable loss of electrons between two adjacent Yb$^{3+}$ ions, implying the existence of the repulsion between adjacent Yb$^{3+}$ ions, which is also revealed in the calculations of the pair interaction energy ($\Delta E^{(n=2)}$) as shown in Fig. \ref{pair-fig} (b). For wurtzite structure, any two nearby substitutional Yb ions on the $ab$ basal plane experience elastic repulsion due to the fact that (1) the repulsion exists between them, and (2) the Al ions around them have to relax away from the closest Yb ions (Yb$^{3+}$ has a larger ionic size), and hence the total energy increases, i.e. repulsion. In contrast, Fig. \ref{CHG} (b) displays that a deformed bipyramidal structure can be formed in the certain local chemical environment, i.e. a chain configuration (--Yb--N--Yb--) along $c$ axis, where charges are accumulated on both sides of N ions in the $c$ direction with the bond lengths of 2.200 and 2.464 \AA for Yb--N(1) and Yb--N(2), respectively. Such a chain configuration allows N ion to attract two nearby Yb ions, and causes a decrease in the total energy as revealed from the calculations of pair interaction energy ($\Delta E^{(n=3)}$). Owing to different bond lengths for Yb--N(1) and Yb--N(2), the deformed bipyramidal structure maintains the piezoelectricity along $c$ axis.

\begin{figure}[ht]
\begin{center}
\includegraphics[width=8 cm]{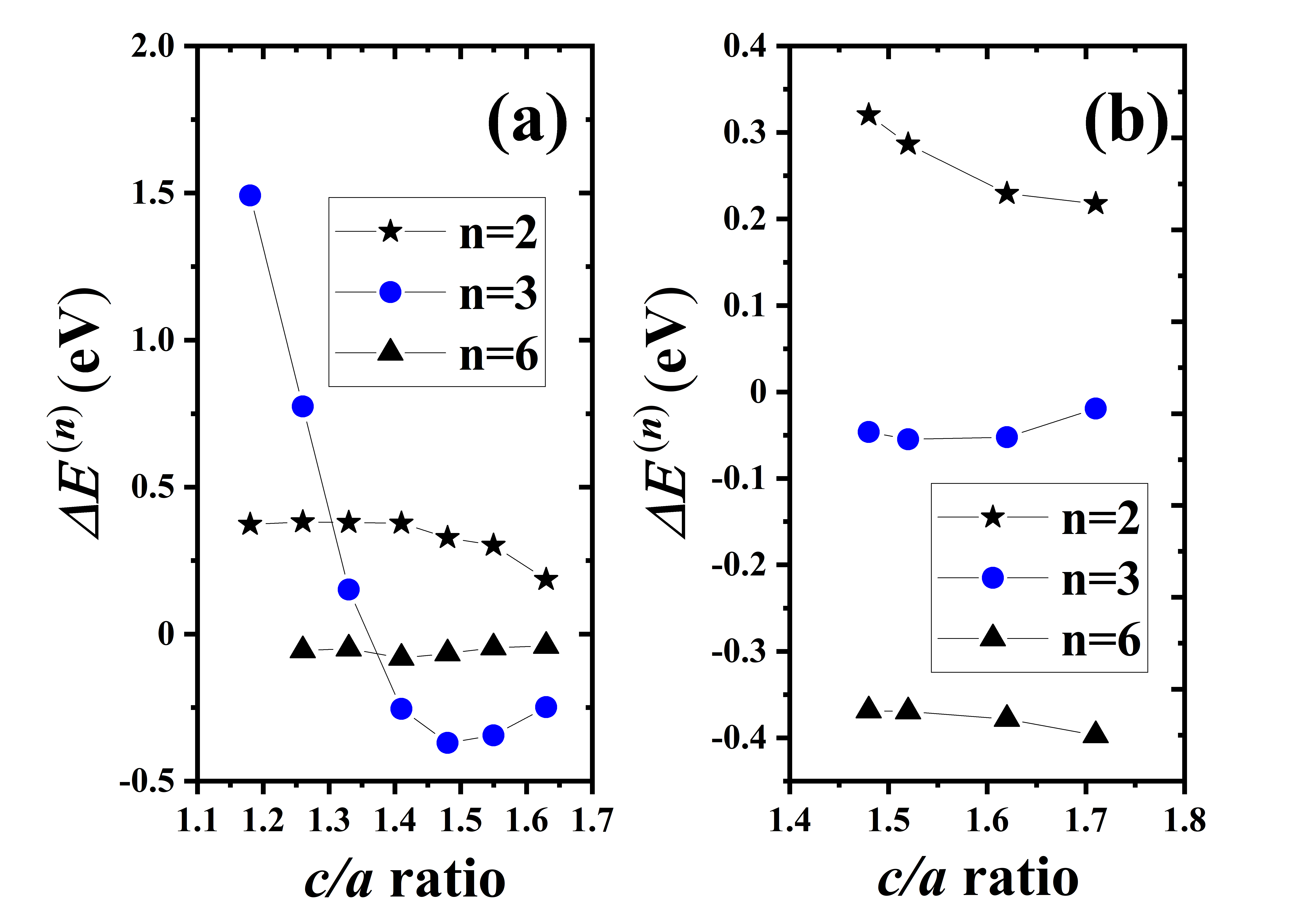}
\caption{Calculated pair interaction energies $\Delta E^{(n)}$ at different $c/a$ ratios: (a) the lattice spacing in $a$ axis is fixed, and (b) the lattice spacing in $c$ axis is fixed. $\Delta E^{(3)}$ is the pair interaction energy due to the neighbor pair along $c$ axis, and $\Delta E^{(2)}$ and $\Delta E^{(6)}$ are the pair interaction energies due to the neighbor pairs on the $ab$ basal plane in wurtzite (Yb,Al)N. Here, a 3 $\times$ 3 $\times$ 2 AlN structure with the lattice parameters of (Yb,Al)N ($x$=0.33) was selected as the reference, and a 3 $\times$ 3 $\times$ 3 $k$--mesh was used for all the calculations.}
\label{pair2}
\end{center}
\end{figure}

In Table \ref{cij}, $C_{11}$ and $C_{33}$ rapidly decreases with increasing $x$, indicating structural flexibility along $a$--axis and $c$--axis direction for (Yb,Al)N under the external uniaxial stress, respectively. To understand the origin of structural flexibility at $x$=0.33, we calculated the pair interaction energy $\Delta E^{(n)}$ under different $c/a$ ratios. Fig. \ref{pair2} (a) displays that $\Delta E^{(3)}$ keeps negative down to $c/a$=1.35, and then becomes positive, indicating the elastic nature of the pair interactions. When a Yb atom is substituted into Al site, a Yb--N(1) chemical bond is formed along $c$ axis, and gives rise to the decrease in the total energy of the system, which has been demonstrated by our calculations (not shown here). A chain configuration of --Yb--N--Yb-- can further reduce the total energy. Because the lattice spacing in $a$ axis was fixed in the calculations, $\Delta E^{(2)}$ and $\Delta E^{(6)}$ almost remain the same positive values at all c/a ratios.

On the other hand, Fig. \ref{pair2} (b) displays that a strong repulsive interaction exists between Yb--Yb pair along $a$ axis because $\Delta E^{(2)}$ keeps a large positive value for all $c/a$ ratios with a fixed $c$ lattice spacing. For wurtzite structure ($u$$\neq$0.5), two nearby substitutional Yb ions on the basal plane experience elastic repulsion because (1) Yb$^{3+}$ has a larger ionic size than Al$^{3+}$ and (2) N ions are not on the same basal plane with Yb ions. Overall, because of the elastic nature of the attraction or the repulsion as shown in Fig. \ref{pair2}, the Yb--Yb pair interactions are considered a cause of structural flexibility in (Yb,Al)N. 

\section{Conclusion}

Our experiments show that alloying cubic YbN into wurtzite AlN leads to enhanced electromechanical coupling in AlN, where $k^2_t$ is improved by 140\% at $x\sim$11\%. Theoretical calculations agree well with experimentally measured $k^2_t$ and elastic constants, suggesting that electromechanical coupling is mainly dominated by the local strain around Yb ions and structural flexibility caused by Yb substitution into host Al sites. Introducing more YbN into the AlN system induces mechanical softening up to $x$=43\%. Our calculations indicate that the Yb--Yb pair interactions can contribute to such mechanical softening. The calculations of pair interaction energies also show that the Yb-Yb pairing avoids $n$=1, 2, and 5 nearest neighbor configurations, and favors $n$=3, 4, and 6 neighbor configurations, especially $n$=3, i.e. the Yb--Yb pair along $c$ axis. Alloying AlN with YbN is promising, and offers an increased $k^2_t$ up to $\sim$10\%, even at a rather low Yb concentration ($\sim$11\%). Such enhancement in $k^2_t$ is comparable with that of commercial (Sc,Al)N films. 

\section{ACKNOWLEDGMENTS}

T. Yanagitani thanks the support from JST CREST (Grant No. JPMJCR20Q1), Japan. J. Jia acknowledges the funding from JSPS KAKENHI Grant-in-Aid for Scientific Research (C) (Grant No. 20K05368).

\end{document}